\documentclass[prl,showpacs,amsmath,amssymb,twocolumn]{revtex4-1}
\usepackage{graphicx}
\usepackage{color}
\usepackage[usenames,dvipsnames]{xcolor}
\usepackage{natbib}
 
\usepackage{soul}
\usepackage{lmodern} 
\usepackage{placeins}

\newcommand{\mean}[1]{\ensuremath{\left\langle #1 \right\rangle}}

\newcommand{\ad}{\ensuremath{a^\dagger}}
\newcommand{\bd}{\ensuremath{b^\dagger}}

\newcommand{\XM}{\ensuremath{X_\textrm{M}}}
\newcommand{\PM}{\ensuremath{P_\textrm{M}}}
\newcommand{\XL}{\ensuremath{X_\textrm{L}}}
\newcommand{\PL}{\ensuremath{P_\textrm{L}}}
\newcommand{\XMO}{\ensuremath{X_\textrm{M,out}}}
\newcommand{\PMO}{\ensuremath{P_\textrm{M,out}}}
\newcommand{\XLO}{\ensuremath{X_\textrm{L,out}}}
\newcommand{\PLO}{\ensuremath{P_\textrm{L,out}}}

\newcommand{\Nbar}{\ensuremath{\bar{N}}}
\newcommand{\nbar}{\ensuremath{\bar{n}}}

\newcommand{\omegam}{\ensuremath{\omega_{\textrm{M}}}}

\newcommand{\Sq}[1]{\ensuremath{\left [#1\right]}}

\newcommand{\kb}{\ensuremath{k_\textrm{B}}}

\newcommand{\myref}[1]{(\ref{#1})}

\begin{document}

\title{Quantum optomechanics beyond the quantum coherent oscillation regime}

\author{Kiran Khosla$^1$}
\email{k.khosla@uq.edu.au}
\author{George A. Brawley$^1$}
\author{Michael R. Vanner$^{1,2}$}
\author{Warwick P. Bowen$^1$}
\affiliation{$^1$Center for Engineered Quantum Systems, University of Queensland,
	St Lucia 4072, Australia}
\affiliation{$^2$Clarendon Laboratory, Department of Physics, University of Oxford, OX1 3PU, United Kingdom}


\begin{abstract}
Interaction with a thermal environment decoheres the quantum state of a mechanical oscillator. When the interaction is sufficiently strong, such that more than one thermal phonon is introduced within a period of oscillation, quantum coherent oscillations are prevented. This  is generally thought to preclude a wide range of quantum protocols. Here, we introduce a pulsed optomechanical protocol that allows ground state cooling, general linear  quantum non-demolition measurements, optomechanical state swaps, and quantum state preparation and tomography without requiring quantum coherent oscillations. Finally we show how the protocol can break the usual thermal limit for sensing of impulse forces. 
\end{abstract}
\maketitle

\textit{Introduction} --- Quantum optomechanics uses an optical or microwave field  to prepare, control and characterize the quantum states of a meso --- to macroscopic mechanical oscillator, typically using a cavity to enhance the interaction~\cite{milburn_introduction_2011,kippenberg_cavity_2007,anetsberger_near-field_2009,eichenfield_optomechanical_2009,thompson_strong_2008,sankey_strong_2010}. Optomechanical systems have been proposed for quantum information applications~\cite{stannigel_optomechanical_2011,mcgee_mechanical_2013,stannigel_optomechanical_2012}, tests of foundational physics~\cite{pikovski_probing_2012,marshall_towards_2003}, and are currently used for state of the art sensors~\cite{gavartin_hybrid_2012,xu_squeezing_2014,forstner_cavity_2012}, with each application requiring different optomechanical regimes. The significant progress in devices and technology of the last decade~\cite{aspelmeyer_cavity_2014} suggest some of these associational targets will be realisable in the near future. With the exception of position non-demolition measurements and their derivatives~\cite{vanner_pulsed_2011}, it is generally considered that operation in the quantum coherent oscillation (QCO) regime --- where on average less than one thermal phonon is exchanged per mechanical period, and the oscillator remains coherent for at least a single oscillation --- is a minimum requirement for such experiments~\cite{clerk_back-action_2008,aspelmeyer_cavity_2014,norte_mechanical_2016,meenehan_thermalization_2014,tsaturyan_ultra-coherent_2016}. This notion is reflected in recent theoretical and experimental results~\cite{naik_cooling_2006,purdy_observation_2013,hertzberg_back-action-evading_2010,aspelmeyer_cavity_2014,kronwald_arbitrarily_2013,woolley_two-mode_2013,chan_laser_2011,teufel_sideband_2011,riviere_optomechanical_2011,anetsberger_cavity_2011,marshall_towards_2003}. In the high temperature limit, for a full mechanical period to remain coherent, requires $Q\omegam > \kb T/\hbar$ with $Q$ the quality factor of the mechanical oscillator of frequency \omegam~at temperature $T$. This places stringent constraints on both the temperature and mechanical resonance frequency for which optomechanics protocols can be implemented. Highly desirable room temperature operation requires $\omegam Q \gtrsim 5\times 10^{14}$~\cite{meenehan_thermalization_2014} which is beyond current technology for low frequency oscillators~\cite{aspelmeyer_cavity_2014} relevant to precision force sensing~\cite{gavartin_hybrid_2012} and tests of macroscopic quantum mechanics. Here we show that by applying a pulsed optomechanics protocol, ground state cooling, general linear back-action evading measurements, state swaps, and non-classical state preparation and quantum tomography are all possible outside the QCO regime. Furthermore, when applied in the classical regime, our technique provides a pathway to evade thermal force noise which limits current state of the art force sensors. Such sensors operate in the non-QCO regime using a ``time of flight'' method to translate a force signal --- coupled directly to momentum --- into position which can be optically read out~\cite{caves_quantum-mechanical_1980,xu_squeezing_2014}. The necessary time delay introduces thermal force noise, reducing the sensitivity of the measurement. Implementing the measurement over a small fraction of a period reduces the thermal noise, increasing the measurement sensitivity.

Our results extend previous work on speed meters, which have been proposed for gravitational wave detectors~\cite{braginsky_gravitational_1990,graf_design_2014}. Speed meters can achieve quantum non-demolition measurements of the relative momentum of two oscillators. Unlike our approach, they have generally been studied for detecting oscillating forces in a frequency band far from the mechanical resonance~\cite{danilishin_sensitivity_2004,purdue_practical_2002-1,purdue_analysis_2002,chen_sagnac_2003}, and within the free mass approximation.

\textit{Model} --- Pulsed optomechanics~\cite{vanner_pulsed_2011,braginskii1978optimal} generates optomechanical correlations over a short time scale compared to free mechanical evolution. These short interactions can be used to manipulate the state of a mechanical oscillator, greatly reducing mechanical thermalisation for a given protocol. The interaction Hamiltonian for such a system is given by $H_{\mathrm{I}} = \hbar g_0 \ad a (b + \bd)$, where $a$ ($b$) is the annihilation operator for the optical (mechanical) mode and $g_0$ is the bare optomechanical interaction rate. We restrict the analysis to the linearized interaction where the optical field is linearised, $a \rightarrow \alpha + a$, about a time dependent amplitude $\alpha(t) = \mean{a(t)}$ where, without loss of generality we define $\alpha$ to be real~\cite{Bowen_Optomechanics}. The pulse envelope $\alpha(t)$ is chosen to be a Gaussian with pulse width $\tau$. Expanding to first order in $a$, the interaction Hamiltonian is given by $H_{\mathrm{I}} / \hbar =  g_0 \alpha( a + \ad)(b + \bd) - g_0\alpha^2(b + \bd)$. The first term is the linearised optomechanical interaction, and the second term is a coherent momentum displacement. This displacement is deterministic and can be cancelled by applying an opposite classical displacement to the oscillator; we therefore neglect it henceforth. Outside the single photon strong coupling regime the second order term, $g_0 \ad a (b +\bd)$ that has also been neglected, remains negligibly small, even when the envelope $\alpha(t) \rightarrow 0$, as the mechanical dynamics is dominated by Brownian motion~\cite{khosla_quantum_2013}. The cavity is modeled as a single sided cavity with decay rate $\kappa$ which is large enough for the optical pulse to adiabatically interact with the cavity $\tau  \gg 1/\kappa$. In this regime the intracavity field is proportional to the input field $\alpha(t) = \alpha_{\mathrm{in}}(t)\sqrt{2/\kappa}$ where the input field is normalized $\int |\alpha_{\mathrm{in}}|^2 dt = \Nbar$ where \Nbar~is the mean photon number in the pulse. Furthermore, as in other pulsed optomechanics protocols~\cite{vanner_pulsed_2011,khosla_quantum_2013,bennett_quantum_2016}, the mechanical oscillator is assumed frozen during a single pulsed interaction, so that $\tau \ll 1/\omegam$ with the optomechanical system necessarily operating in the unresolved sideband limit, $\kappa \gg \omegam$ (bad cavity limit).

Under these conditions the unitary describing the total pulsed interaction~\cite{vanner_pulsed_2011} is given by $U(\XM,\XL) = \exp [-i\lambda \XM \XL ]$, where $\sqrt{2}\XM =b + \bd $ and $\sqrt{2}\XL = a + \ad$ are the mechanical position and optical amplitude quadratures, respectively. The dimensionless constant $\lambda = 4(2\pi)^{1/4}\sqrt{\tau \Nbar g_0^2/ \kappa}$ is the optomechanical interaction strength~\cite{vanner_pulsed_2011}. The unitary correlates the optical phase quadrature ($\PL$) and mechanical position as $U^\dagger\PL U = \PL - \lambda \XM$ at the expense of adding back action to the momentum ($\PM$) of the oscillator $U^\dagger\PM U = \PM - \lambda \XL$. We will now show that a sequence of such pulsed interactions, allows arbitrary mechanical quadrature measurements which are sufficient for quantum state tomography, squeezed state preparation, and pulsed optomechanical state swaps~\cite{bennett_quantum_2016}. Outside the QCO regime, a single pulsed measurement cannot achieve these important goals.

\begin{figure}
	\centering
	\includegraphics[width=\columnwidth]{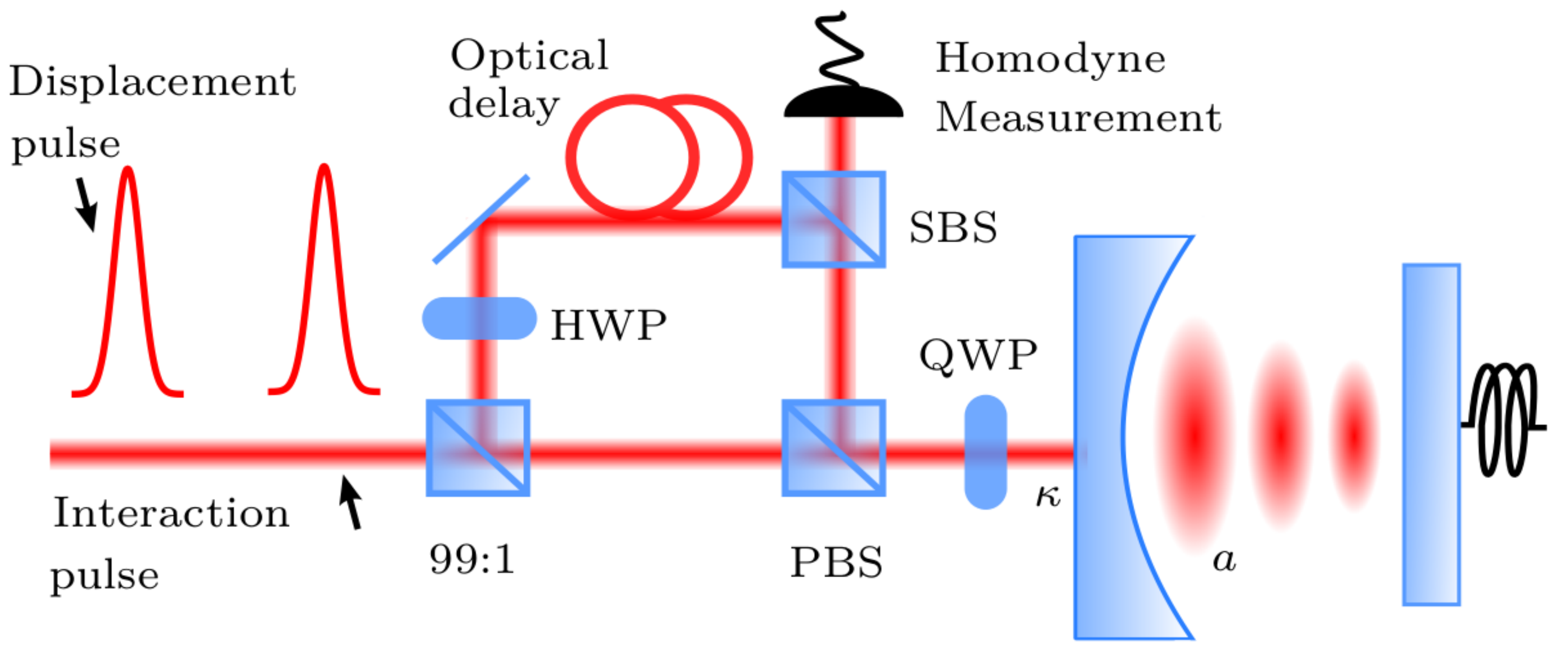}
	\caption{\small Schematic of the protocol. A pulse interacts twice with a mechanical element inside an optical cavity. A displacement between pulses used to change the coherent amplitude, and thus the second optomechanical interaction strength. After the second interaction a switchable beam splitter (SBS) sends the pulse to be measured via homodyne detection. Half and quarter wave plates (HWP and QWP respectively) and a polarizing beam splitter (PBS) are used to direct the optical pulse.}%
	\label{fig:sch}%
\end{figure}

In our back action evading protocol a single optical pulse interacts with the mechanical element twice (figure~\ref{fig:sch}). This process is represented by an initial optomechanical interaction, free mechanical evolution $\theta = \omegam t$, and finally the second optomechanical interaction with the same optical field but different interaction strength. Neglecting decoherence (which will be included in the next section), the protocol is described by the overall unitary
\begin{eqnarray}
\mathcal{U} &=& \exp[ - i\lambda_2 \XM \XL]e^{-i\theta  \bd b} \exp[ -i\lambda_1 \XM \XL] \\
&=& R(\theta) ~\chi(\lambda_1\lambda_2) ~U(\XL\XM^\phi)
\label{eq:unit}
\end{eqnarray}
This unitary, acting from right to left, can be understood as follows. The first term, $U(\XL\XM^\phi) = \exp(-i\mathcal{G}\XL\XM^\phi)$, is the optomechanical interaction between a rotated mechanical quadrature $\XM^\phi = \XM \cos\phi + \PM \sin\phi$ and the amplitude quadrature, $\XL$ of the light with (dimensionless) interaction strength $\mathcal{G} = (\lambda_1^2 + \lambda_2^2 + 2 \lambda_1\lambda_2\cos\theta)^{1/2}$, and rotated quadrature angle 
$\tan\phi= \lambda_2\sin\theta (\lambda_2 \cos\theta + \lambda_1)^{-1}$. We therefore see that by applying two interactions it is possible to generate a single effective $\XL\XM^\phi$ interaction for an arbitrary mechanical quadrature $\phi$. For a given $\phi$, the natural choice of free parameters ($\lambda_1,~\lambda_2,\theta$) are the set that maximizes $\mathcal{G}$. The second unitary in Eq~(\ref{eq:unit}), $\chi(\lambda_1\lambda_2) = \exp[\frac{i}{2} \lambda_1\lambda_2\XL^2 \sin\theta]$ is an optical Kerr nonlinearity, the pulsed analogue of the effects reported in Refs~\cite{purdy_strong_2013,safavi-naeini_squeezed_2013}. The final term $R(\theta) = e^{-i\theta \bd b}$ is the mechanical rotation due to the delay between pulses, and must be accounted for in the final mechanical state. Different interaction strengths $\lambda_1$ and $\lambda_2$, with positive or negative sign, may be chosen by using a displacement pulse to change the mean photon number in-between the two interactions. The total unitary couples the optical and mechanical states as 
\begin{subequations}
\begin{eqnarray}
\XM &\rightarrow& \XM\cos\theta + \PM\sin\theta - \XL\lambda_1 \sin\theta \\
\PM &\rightarrow& \PM\cos\theta - \XM \sin\theta - \XL\Sq{\lambda_2 - \lambda_1\cos\theta} \\
\XL &\rightarrow& \XL \\
\PL &\rightarrow&  \PL -\mathcal{G}\XM^\phi + \XL  \lambda_2\lambda_1 \sin\theta. 
\end{eqnarray}
\label{eq:4}
\end{subequations}
\noindent~For a single optomechanical interaction ($\lambda_2=\theta  = 0$) the back action noise ($\XL$) is necessarily imparted onto the momentum of the oscillator. However, if the interaction strengths are chosen such that $\lambda_2 = -\lambda_1 \cos\theta$ the back action on the momentum exactly cancels and a homodyne measurement of the optical field can be used to conditionally prepare a momentum squeezed state. This can be done in an arbitrarily short time, $\theta \rightarrow 0$ at the expense of increasing the overall interaction strength $\mathcal{G}$. It can therefore introduce arbitrarily low levels of the thermal noise that precludes momentum squeezing with other protocols outside the QCO regime. By varying the interaction strengths for the two pulses $(\lambda_1, \lambda_2)$, one may deterministically choose which mechanical quadrature the back-action is added to, and which quadrature is back-action free. 

For the unitary case, or indeed, in the QCO regime where a negligible amount of thermal noise is introduced after a single oscillation, there is no need to implement the two pulse protocol. A momentum measurement can be achieved with $\theta = \pi/2$, $\lambda_1 = 0$ (wait then measure), correlating the phase quadrature with the initial momentum of the oscillator. Alternatively, momentum state preparation can be achieved with $\theta = \pi/2, \lambda_2 = 0$ (measurement then wait), preparing a conditional position squeezed state that rotates into momentum a quarter cycle later. However, outside the QCO regime phonon exchange during the necessary $\theta = \pi/2$ delay thermalizes the mechanical state, resulting in additional measurement noise and degrading conditional state preparation, even in the limit of arbitrarily large optomechanical interaction strengths. Equations~(\ref{eq:4}a)-(\ref{eq:4}d) illustrate how, with our protocol, one can reduce the duration of the protocol $\theta$ ---  and thus the thermalisation --- while still generating the optomechanical correlations necessary to prepare and measure an arbitrary mechanical quadrature.

\textit{Nonunitary evolution}\label{sec:nonunitary} --- In this section, the description of the protocol is extended to include mechanical dissipation and optical losses~\cite{caldeira_path_1983,schmidt_c._1985}. During each pulsed interaction, it is still assumed that the mechanical oscillator remains effectively frozen only being free to evolve during the wait time between pulses. The cavity is treated as highly over coupled so that the only significant source of optical loss occurs during the storage time between the two interactions. In this case, each optomechanical interaction is well described by the unitary operator in equation~\myref{eq:unit}. Thermalisation is included during the free evolution of the oscillator by the Langevin equation of motion~\cite{benguria_quantum_1981}
\begin{subequations}
\begin{eqnarray}
\dot \XM &=& \omegam \PM  \label{eq:1}\\
\dot \PM &=& -\omegam \XM - \gamma \PM + \sqrt{2\gamma}\xi \label{eq:2}
\end{eqnarray}\label{eq:3s}
\end{subequations}%

\noindent where $\gamma$ is the oscillator decay rate and $\xi(t)$ is a zero mean white noise operator with correlation function $\mean{\xi(t) \xi (t')} = (\nbar + \frac12) \delta(t-t')$ where $\nbar \approx \kb T / \hbar \omegam$ in the high temperature limit. The thermal force satisfies $[\XM(t), \xi(t)] = \sqrt{\gamma/2}i$ to preserve the commutation relations. Optical losses are modeled as a single beam splitter unitary $\mathcal{B}_\eta$ with intensity loss $\eta$ after each optomechanical interaction. The protocol is then given by $\mathcal{B}_{\eta} U_2 \mathcal{B}_{\eta} \mathcal{M}_{\theta}U_1$, where $U_i$ is the $i$th optomechanical unitary, and $\mathcal{M}_\theta$ is a non-unitary map that describes the dissipative mechanical evolution given by Eqs~(\ref{eq:1}) and~(\ref{eq:2}) over the rotation angle $\theta$. After the full protocol the optical and mechanical quadratures are given by,

\begin{subequations}
\begin{eqnarray}
\XMO &=& \XM\cos\theta + \PM\sin\theta + \xi_X - \lambda_1\XL\sin\theta \label{eq:6a}\\
\PMO&=& \PM\cos\theta -\XM\sin\theta + \xi_P - \nonumber\\
& & \XL(\sqrt{\eta} \lambda_2 - \lambda_1\cos\theta ) - \lambda_2\sqrt{1-\eta}\delta X_1 \\
\XLO &=& \eta\XL + \sqrt{\eta - \eta^2}\delta X_1 + \sqrt{1 - \eta}\delta X_2 \\
\PLO &=& \eta\PL + \sqrt{\eta - \eta^2}\delta P_1 + \sqrt{1 - \eta}\delta P_2 \nonumber \\
& &-\sqrt{\eta}\lambda_1\lambda_2\sin\theta \XL + \mathcal{G}\XM^\phi -\sqrt{\eta}\lambda_2\xi_X \label{eq:6d}
\end{eqnarray}
\end{subequations}

\noindent where $\xi_{X(P)}$ is the thermal noise added to the mechanical position (momentum) during the non-unitary evolution and $\delta X_i$~($\delta P_i$) is the amplitude (phase) quadrature vacuum noise entering from the $i$th optical loss channel. With decoherence included, the new measured quadrature and measurement strength are given by $\phi = \arctan [\lambda_2\sin\theta(\lambda_2\cos\theta + \sqrt{\eta}\lambda_1)^{-1}]$ and $\mathcal{G} = (\eta^2\lambda_1^2 + \eta\lambda_2^2 + 2\lambda_1\lambda_2\eta^{3/2}\cos\theta)^{1/2}$ respectively. Eq~(\ref{eq:6d}) highlights how the scheme works in the presence of mechanical thermalisation. A measurement of $\PLO$ provides information about $\mathcal{G}\XM^\phi$ with all other terms contributing zero mean Gaussian noise. For small $\theta$ the thermal noise in the position increases as $\theta^3$ scaling (see Ref.~\cite{bennett_quantum_2016} for details). Consequently, reducing the duration of the protocol, cubicly reduces this noise term, which is generally the dominant noise source outside the QCO regime where $\nbar/Q \gg 1$. Due to correlations between the phase and amplitude quadratures, i.e. the Kerr term $\sqrt{\eta_2}\lambda_1\lambda_2\sin\theta\XL$ in Eq~(\ref{eq:6d}), extra noise is added to the phase quadrature. Since this noise is correlated to the optical amplitude quadrature, $\XL$, it can be reduced by measuring a rotated optical quadrature $\XL^\varphi = \XL\cos\varphi + \PL\sin\varphi$ where the optimal angle $\varphi$ is determined numerically. In the following section, we show how the correlations in Eqs~(\ref{eq:6a})-(\ref{eq:6d}) can be used for state preparation, measurement and force sensing.

\textit{Measurement and tomography} --- For quantum measurement and tomography, the aim is to measure the statistics of the \textit{a priori} mechanical state, $\XM^{\phi}$ for each $\phi \in [0,\pi)$, without making any assumptions on the statistics of $\XM^\phi$. From Eq~(\ref{eq:6d}), a measurement of $\PLO$ (or indeed any optical quadrature $\XL^\varphi \neq \XL$) is a measurement of $\XM^\phi$ with all other terms contributing zero mean Gaussian noise. The uncertainty in a Gaussian variable $A$, given a measurement of $B$ is given in general by the conditional variance, $V(A|B) = V(A) - C(A,B)^2/V(B)$, where $C(A,B) \equiv \frac12 \langle AB+BA\rangle - \langle A\rangle \langle B\rangle$ is the correlation between $A$ and $B$. When the conditional variance of the measured quadrature $V(\XM^\phi|\XL^\varphi)$ is below the ground state variance (of 1/2), full quantum state tomography can be performed efficiently, and Wigner negativity can be directly observed~\cite{lvovsky_quantum_2001,vanner_towards_2015}.  Outside the QCO regime direct observation of Wigner negativity is not possible, using either single pulse, or stroboscopic measurement; while state tomography becomes increasingly challenging due to the convolution of the thermal noise in the measurement results. Both can be achieved, in principle, to an arbitrary degree of accuracy here. To the authors knowledge, this is the first protocol capable of efficient state tomography for oscillators outside the QCO regime. 

\begin{figure}%
\centering
\includegraphics[width=\columnwidth]{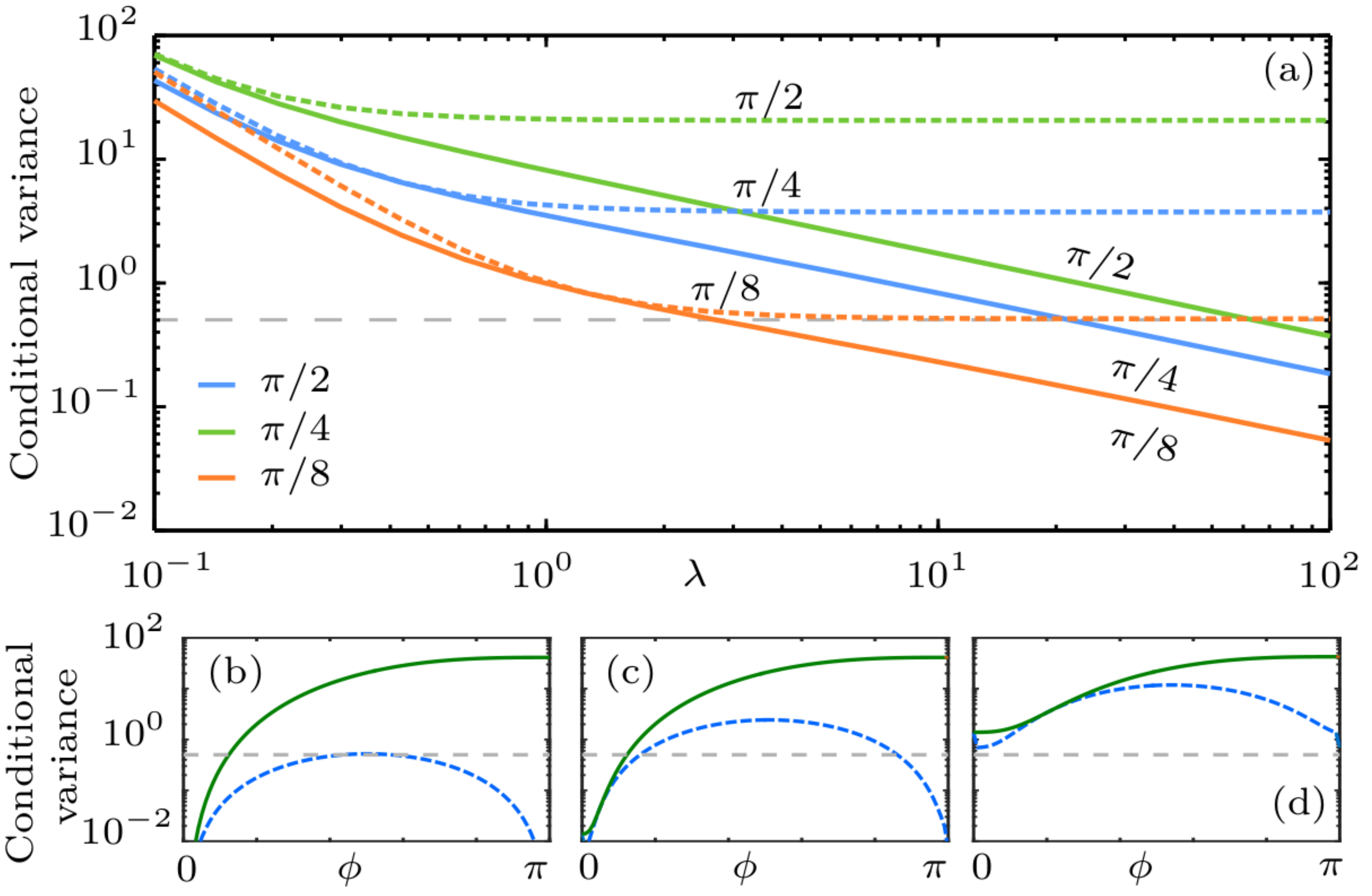}
\caption{ \small (a) Conditional variance for measurement of different quadratures ($\phi = \pi/2,\pi/4,\pi/8$), using the double (solid lines) or single (dashed lines) interaction protocols. (b)-(d) Conditional variance as a function of mechanical quadrature angle for the double (dashed lines) and single (solid lines) interaction protocols for $\lambda = 60$, $\lambda = 6$ and $\lambda = 0.6$ respectively, and bath temperature of 100 K. Grey dashed lines indicate the ground state variance and $\omegam/2\pi = 1$ kHz, $\gamma/2\pi = 1$ Hz.}%
\label{fig:snr}%
\end{figure}

Figure~\ref{fig:snr} compares the two approaches for realistic experimental parameters. The parameters are chosen as $\omegam/2\pi = 100$ kHz, $\gamma/2\pi =1$ Hz ($Q = 10^5$), similar to the silicon carbide resonators in Ref~\cite{kermany_factors_2016}. At a temperature of 1 K, $2\pi\nbar/Q \approx 82$ phonons enter per oscillation, residing well outside of the QCO regime. The bare optomechanical coupling rate is conservatively set at $g_0/2\pi = 1$ Hz with a cavity decay rate $\kappa/2\pi =1$ GHz (optical $Q \approx 2.5\times 10^5$ at 1064 nm). To ensure a fair comparison, we choose to relate the single pulse interaction strength $\lambda$ to the two pulse interaction strengths via $\lambda = \sqrt{\lambda_1^2 + \lambda_2^2}$ such that the mean photon number in the single pulse is equal to the sum of mean photon numbers in the two pulse scheme. As shown in figure~(\ref{fig:snr}a), there is lower bound for the conditional variance using a single pulsed interaction, independent of the interaction strength. This lower bound is removed using our protocol, allowing sub-ground state resolution outside the QCO regime. Figures~(\ref{fig:snr}b)-(\ref{fig:snr}d) shows that sub-ground state resolution is possible for all quadrature angles allowing for full quantum state tomography for $\lambda \gtrsim 60$.

\begin{figure}%
\centering
\includegraphics[width=\columnwidth]{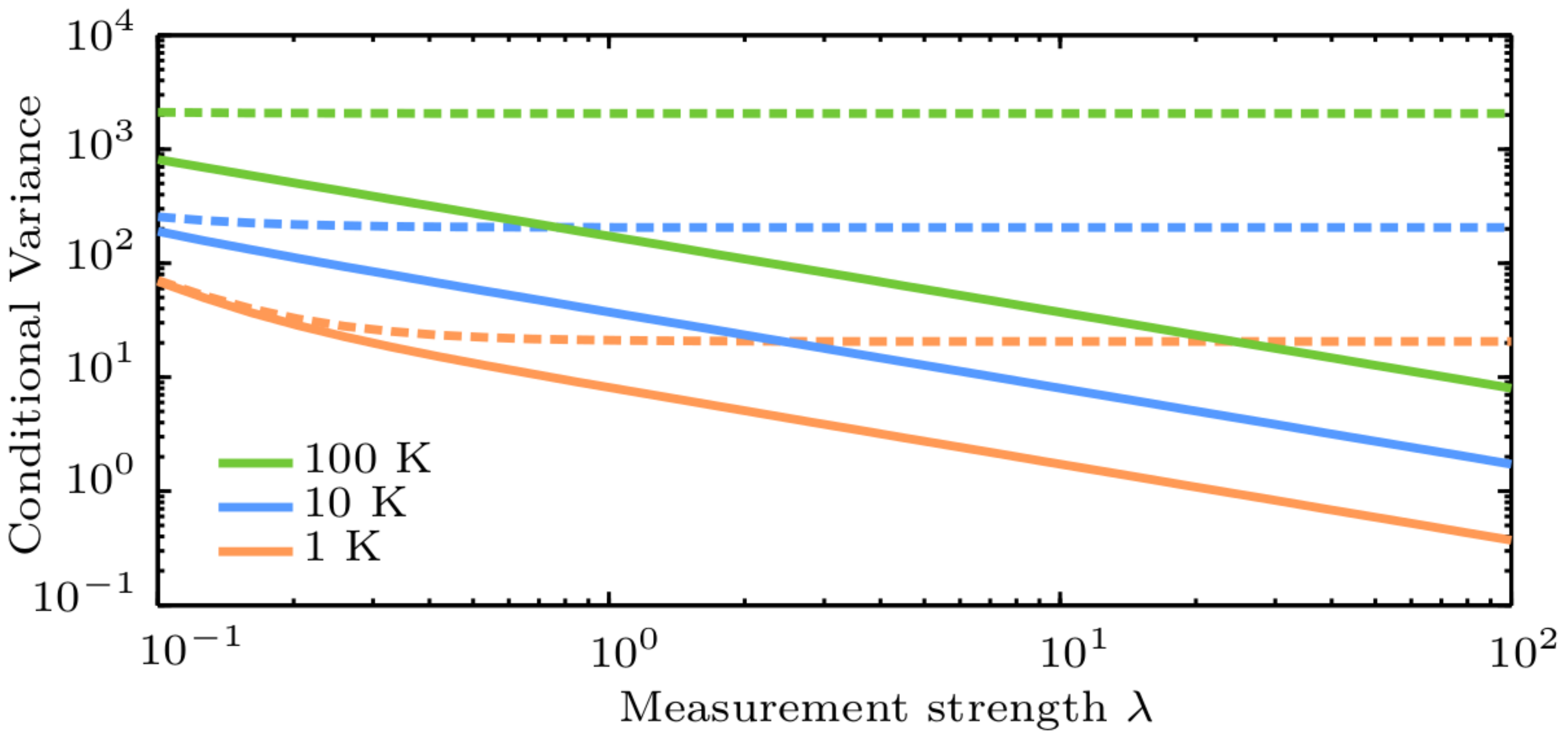}
\caption{\small Conditional variance of the final oscillator momentum using the double (solid lines) and single (dashed lines) interaction protocols at different temperatures. A temperature of 100 K corresponds to 8200 phonons exchanged per cycle for $\omegam/2\pi = 1$ kHz, $\gamma/2\pi = 1$ Hz.}%
\label{fig:StatePrep}%
\end{figure}

\textit{Squeezed state preparation} --- We now turn to how the protocol can be used to prepare a squeezed mechanical state in the presence of thermalization. Mechanical squeezing such as ponderomotive squeezing~\cite{fabre_quantum-noise_1994} or reservoir engineering~\cite{jahne_cavity-assisted_2009} requires many oscillations of the mechanical oscillator and therefore cannot be implemented within the QCO regime. A single pulsed interaction can be used to generate squeezing outside the QCO regime but is best suited to position squeezing~\cite{vanner_pulsed_2011}. Our scheme allows squeezing of an arbitrary mechanical quadrature outside the QCO regime. Due to the finite rotation during the protocol there is a subtle distinction between squeezed state preparation and quantum measurement. For state preparation the aim is to condition the variance of the \textit{a posteriori} quantum state, V$(\XMO^\phi|\PL)$ instead of the \textit{a priori} state, V$(\XM^\phi|\PL)$. Figure~(\ref{fig:StatePrep}a) shows the \textit{a posteriori} conditional variance for a momentum measurement at different temperatures, comparing the double interaction protocol introduced here with a single pulsed interaction. It shows arbitrary quadrature squeezing is in principle possible if the interaction strength is high enough, and is necessarily achievable within a fraction of an oscillation. Our scheme can achieve significantly lower variance at higher temperatures, even for small-moderate interaction strengths. As with quantum tomography, our scheme is able to prepare states with variance below the thermalization line that bounds the single pulsed scheme, meaning the conditional variance can be arbitrarily reduced by increasing the interaction strength. This is due to the finite time taken for the position squeezed state to rotate to a momentum squeezed state. From Eqs~(\ref{eq:6a})-(\ref{eq:6d}), it can be seen that the optimal measured quadrature is not orthogonal to the back action quadrature, as a result any squeezing will be accompanied by anti-squeezing larger than the lower bound set by the Heisenberg limit. Combined with optical losses and the mechanical thermalization, this effect reduces the purity of the final mechanical state.

Note that direct measurement and manipulation of the momentum of an oscillator over a short time scale ($\theta/\omegam$) may be used to directly measure impulse forces --- which couple directly to momentum through $dp = F dt$ --- at the thermal limit. Using our pulsed protocol, the momentum state preparation (measurement) can be made immediately before (after) the impulse force, therefore, the only thermal force in the signal is that which enters during the duration of the impulse. In contrast, a position measurement must necessarily wait an additional time period while the force evolves into a displacement, during which, extra thermal noise is added to the signal. For example taking the detectable momentum change to be on the order of the conditional momentum standard deviation of $\sqrt{200}\sqrt{\hbar m \omega/2}$ at 100 K (from figure~\ref{fig:StatePrep} with $\lambda = 1$), corresponds to a thermally limited impulsed force sensitivity of 8 pN over $\frac{1}{50}$$^{\mathrm{th}}$ of an oscillation period (200 ns), compared with 30 pN over $1/4^{\mathrm{th}}$ of an oscillation period (2.5 $\mu$s) for the single pulsed protocol. Finally, we note that when applied to the pulsed state swap proposal of Ref~\cite{bennett_quantum_2016}, or approach also allows state swaps between light and mechanics outside of the QCO regime.
\normalsize

\textit{Summary} --- We have introduced a protocol to realize quantum optomechanics beyond the QCO regime, where quantum state preparation and direct tomography are possible within a fraction of a mechanical period. This enables a mechanism to observe quantum effects in low frequency, high temperature or high mass mechanical systems that would otherwise be masked by thermalisation over a single mechanical period. This research was jointly funded by Australian Research Council Centre of Excellence for Engineered Quantum Systems (EQuS CE110001013), Discovery Project (DP140100734), Future Fellowship (W. P. B., FT140100650) and the Engineering and Physical Science Research Council (EP/N014995/1). 


\bibliography{Bib}{}

\end{document}